\documentstyle[preprint,prl,aps,epsf]{revtex}

\begin{document}

\title{Exact solution of the one-dimensional ballistic aggregation}
\author{L. Frachebourg}
\address{Institut de Physique Th\'eorique}
\address{Ecole Polytechnique F\'ed\'erale de Lausanne}
\address{CH-1015 Lausanne, Switzerland}

\maketitle

\begin{abstract}
 
An exact expression for the mass distribution $\rho(M,t)$ 
of the ballistic aggregation model in one dimension is derived 
in the long time regime. 
It is shown that it obeys scaling 
$\rho(M,t)=t^{-4/3}F(M/t^{2/3})$ with a scaling function 
$F(z)\sim z^{-1/2}$ for $z\ll 1$ and $F(z)\sim \exp(-z^3/12)$ for 
$z\gg 1$. 
Relevance of these results to Burgers turbulence is discussed.

\end{abstract}

\vskip 1truecm

Ballistic aggregation provides a simple model of 
nonequilibrium statistical physics which 
is a natural version of a dissipative 
gas of hard spheres where particles follow the 
basic laws of mechanics.
It consists in a one-dimensional 
gas of point-like massive particles 
which move 
freely until they collide. 
The perfectly inelastic collision of two 
masses conserves the total mass and momentum, while 
dissipation occurs as kinetic energy is loss in each 
collision. 
One can anticipate the formation of more and more massive while slower 
and slower aggregates. 

This model was introduced by Carnevale, Pomeau and Young \cite{cpy} where 
they conjectured, based on scaling arguments and numerical simulations, 
an asymptotic scaling regime for the mass distribution 
$\rho(M,t)=F(M/\langle M\rangle_t)/\langle M\rangle_t^2$. The average mass
per aggregate 
was supposed to grow algebraically with time as $\langle M\rangle_t\sim 
t^{2/3}$ 
and the scaling function had a simple universal 
exponential form $F(z)=\exp(-z)$ independent of the initial conditions.
Later, this conjecture was reinforced by Piasecki \cite{p} where he solved 
the hierarchy of dynamical equations governing the system inside a mean-field
approximation scheme. 

This system, in its continuous limit, was also studied as 
a simplified astronomical 
model for the agglomeration of cosmic dust into macroscopic objects 
\cite{zeld}.
In the ballistic aggregation model, 
the aggregates interact only through their collisions.
An aggregation 
model where gravitational interactions are present has been
studied in \cite{grav}.
  
It is important to mention the connection between 
this model and some solutions 
of the Burgers equation.
At very high Reynolds number, the asymptotic 
solution of the Burgers equation consists of a train of shock waves. 
The laws of motion which govern the dynamics of these shock waves are found
to be equivalent to 
a ballistic aggregation system (see \cite{burgers}).

In this letter, I verify the scaling hypothesis for the mass distribution and 
find in an exact calculation an explicit form for the scaling function.
It happens to be 
different from the conjectured simple exponential.  

Rather than solving the set of partial differential equations governing 
the evolution of the system, I exploit the fact that, 
once the initial state of the system 
is given, the dynamics is completely deterministic. 
Our approach will thus be based on a statistical study of the 
initial conditions and is largely inspired by the work of Martin 
and Piasecki \cite{mp}.

Initially, particles having all the same mass $m$ 
are regularly placed on a line 
with the same inter-particle distance $a$. Initial mass 
density is thus $\rho_0=m/a$.
The initial momentum of the thermalized particles are not correlated and are 
distributed according to the same Gaussian distribution 
$\phi(p)=\sqrt{\beta/(2\pi m)}\exp(-\beta p^2/(2m))$ 
where I now choose $\beta=1/2$ without loss of generality. 

I compute now the density distribution $\rho_m(X,M,P,t)$ where 
$\rho_m(X,M,P,t)dM\,dP\,dX$ is the number 
of aggregates located in $(X,X+dX)$
with momentum in $(P,P+dP)$ and mass in $(M,M+dM)$ at time $t$.
 
When the coordinates $(X,M,P,t)$ of an aggregate are given, 
they uniquely define the number $n=M/m$ as well as the
initial positions
$X-Pt/M-M/(2\rho_0)\leq x_i\leq X-Pt/M+M/(2\rho_0)$ $(i=1,\ldots,n)$
of its constituents.
A crucial point is 
that an aggregate, once formed, is moving according
to the movement of the center of mass (CM) of its constituents, 
which can be determined from the initial state. 
I label the location of the CM at time $t$ of 
the $r$ particles located
initially at $(j+1)a,(j+2)a,\ldots,(j+r)a$ 
by 
\begin{equation}
X_{j+1}^r(t):={1\over rm}\sum_{i=1}^r mx_{j+i}+tp_{j+i}=ja+{r+1\over 2}a+
{t\over rm}\sum_{i=1}^r p_{j+i}.
\label{cm}
\end{equation}

The mass distribution can thus be determined 
from the initial conditions and an aggregate of 
mass $M=mn$ is present at time $t$ iff
\begin{itemize} 
\item{} the CM of its
leftmost $s$ particles has met the CM of its rightmost $n-s$ particles 
for all $s=1,\ldots,n-1$ up to time $t$, leading to 
$X_{j+1}^s(t)>X_{j+s+1}^{n-s}(t)$ for $1\ge s\ge  n-1$,
\item{}
the CM of the successive groups of particles not constituting the aggregate
has not 
met the CM of the aggregate, thus 
$X_{j-r+1}^r(t)<X<X_{j+n+1}^r(t)$ for
$r\ge 1$. . 
\end{itemize}

One has (see \cite{mp} for details)
\begin{eqnarray}
\rho_m(X,M,P,t) &=& \left\langle\prod_{r=1}^\infty
\theta\left\{ X-X_{j-r+1}^r(t)\right\}
\theta\left\{X_{j+r+1}^r(t)-X\right\}\right.\nonumber\cr\\
& & \quad \times \left.
\prod_{s=1}^{n-1} 
\theta\left\{X_{j+1}^s(t)-X_{j+n+1}^{n-s}(t)\right\}
\delta\left(P-\sum_{r=1}^np_{j+r}\right)\right\rangle
\end{eqnarray}
with $\theta$ the Heavyside step function and where
$M=nm$ and $X=(2j+n+1)a/2+tP/M$. 
The brackets denote the 
average over the initial distribution of the momentum. 

Using Eq.(\ref{cm}) and the Gaussian form of the initial distribution, 
one finds the exact 
scaling form
\begin{equation}
\rho_m(X,M,P;t)=\rho_m(M,P;t)={1\over t^{1/3}}\rho_{m'}(M',P')
\end{equation}
with $M'=M/t^{2/3}$, $P'=P/t^{1/3}$ and $m'=m/t^{2/3}$. Note that, due to 
translational invariance, the mass distribution does not depend on $X$.

\begin{figure}
\narrowtext
\epsfxsize=\hsize
\epsfbox{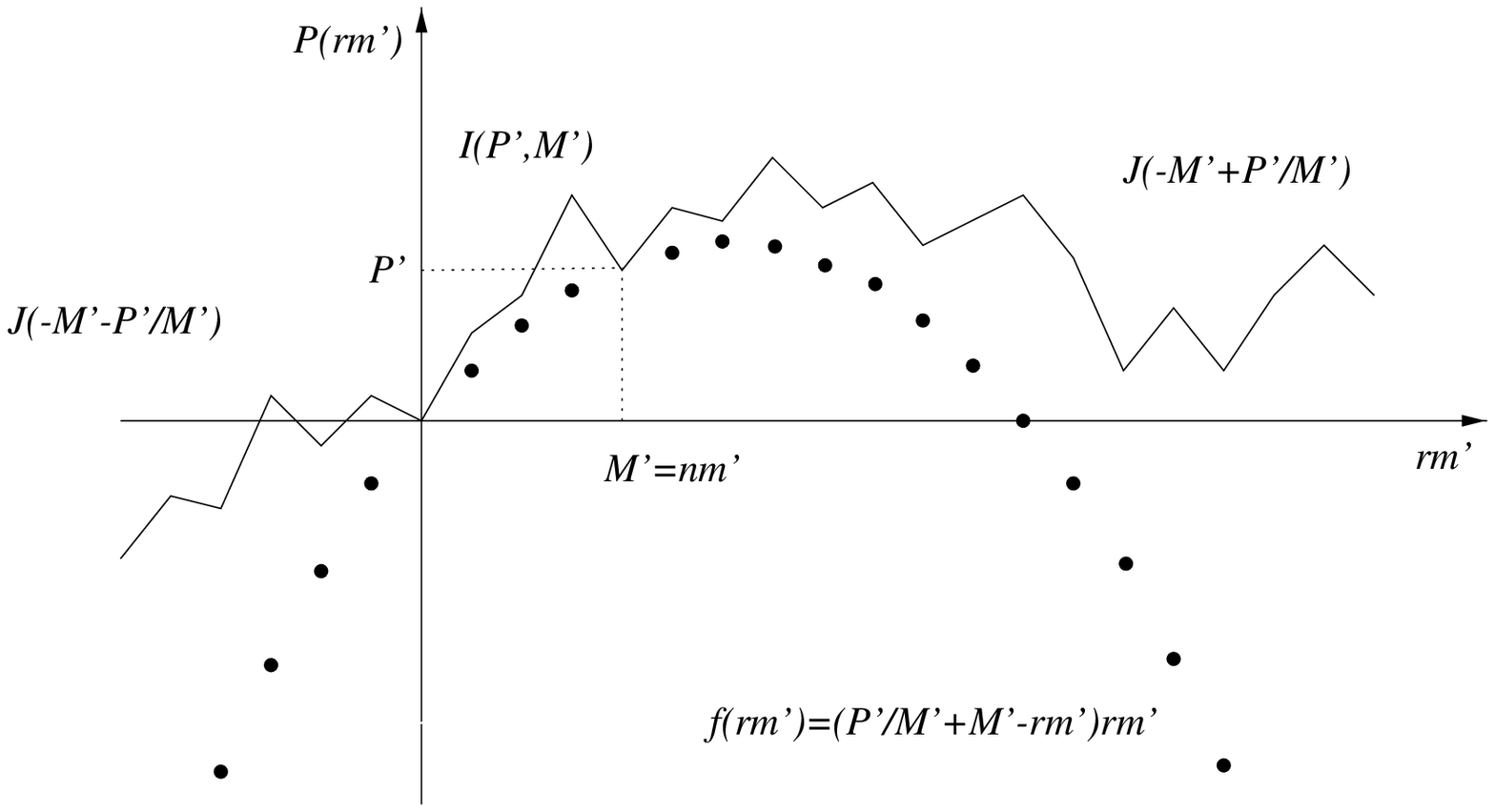}
\caption{
The Brownian motion used in the construction of our solution.
\label{fig1}}
\end{figure}

Owing to the uncorrelated initial Gaussian distribution of 
the momentum, one can compute 
the density $\rho$ using an analogy with a Brownian motion 
in the momentum space 
under particular constraints\cite{mp,fmp}, (see Fig.(\ref{fig1})).
One finds
\begin{equation}
\rho_{m'}(M',P')=J_{m'}\left(-M'-{P'\over M'}\right)I_{m'}(M',P')
J_{m'}\left(-M'+{P'\over M'}\right)
\label{rhoij} \end{equation}
where $J_{m'}(Z)$ is the probability for a Brownian motion $P(\tau)$ 
to start from $P(0)=0$ and pass above the
discrete points $P(rm')>Zrm'-(rm')^2$ ($r\ge 1$), 
and $I_{m'}(M',P')$ is the probability
for a Brownian motion to start at $P(0)=0$, end at $P(M')=P'$ 
and over-passing 
the discrete points $P(rm')>(M'+P'/M')rm'-(rm')^2$ ($1\ge r \ge n$).

I will derive below an expression for the mass distribution in the limit
$m'=m/t^{2/3}\rightarrow 0$ which is reached either 
when $t\rightarrow \infty$
for a fixed $m$ (asymptotic long time limit) or for any fixed time $t$ 
when $m\rightarrow 0$ (continuous limit).
In this limit, one keeps $M'$ and $P'$
of order ${\cal O}(1)$.
In terms of the Brownian motion introduced above, 
the space $m'$ between the discrete points barrier shrinks to zero and 
approaches a continuous barrier which makes the problem tractable 
analytically. 
Nevertheless, the functions $I$ and $J$ are identically null for $m'=0$.
One should thus keeps track of the first 
space $m'$ (the Brownian motion will be unrestricted up to the first 
point of the barrier) 
and will find a mass distribution which is an expansion 
in power of $m'$.
From now on, I drop the subscript $'$ and set $\rho_0=1/2$ without loss 
of generality.

One finds the dominant contribution in $m$:
\begin{eqnarray}
\bar I_m(M,P)&=&{\mathrm e}^{-{P^2\over 2M}}
\int_{Mm-m^2}^{\infty}
dP_1\, \phi(P_1)\int_{-Mm-m^2}^{\infty}
dP_2\,\phi(P_2) K_{M}(P_1,m,P_2,M-m)
\nonumber\\
&=& {m\over \pi}{\mathrm e}^{-{P^2\over 2M}}
\left.{\partial^2\over \partial P_1\partial P_2}K_M(m,P_1,M-m,P_2)
\right|_{P_1=P_2=0}+{\cal O}(m^2)
\label{ik}\end{eqnarray}
and
\begin{eqnarray}
\bar J_m(Y)&=&\int_{Ym-m^2}^{\infty}
dP_1 \phi(P_1)
\lim_{N\rightarrow\infty}
\int_{YNm-(Nm)^2}^{\infty}
dP_2 K_Y(P_1,m,P_2,Nm)\nonumber\\
&=&\sqrt{m\over \pi}\lim_{N\rightarrow\infty}
\int_{YNm-(Nm)^2}^{\infty}
dP_2 \left.{\partial\over \partial P_1} 
K_Y(P_1,m,P_2,Nm)\right|_{P_1=0} +{\cal O}(m^{3/2})
\label{jk}\end{eqnarray}
where $K_Z(P_1,\tau_1,P_2,\tau_2)$ is the probability for a Brownian motion
to start at $P(\tau_1)=P_1$, end at $P(\tau_2)=P_2$ 
while staying above the continuous barrier 
$P(\tau) > f(\tau)=Z\tau-\tau^2$.

Defining the stochastic process $Q(\tau)=P(\tau)-f(\tau)$, one has \cite{mp}
\begin{equation}
K_Z(P_1,\tau_1,P_2,\tau_2)= G(Q_1,\tau_1,Q_2,\tau_2)
\exp\left({1\over 2}(Q_1f'(\tau_1)-Q_2f'(\tau_2)-
{1\over 2}\int_{\tau_1}^{\tau_2}d\tau f'(\tau)^2)\right)
\label{gg}
\end{equation}
where $Q_i=P_i-f(\tau_i)$, $(i=1,2)$. 
The function $G$ is the solution of the equation\cite{mp}
\begin{equation}
\left({\partial \over \partial \tau_2}
-{\partial^2\over \partial Q_2^2}
-{Q_2\over 2}f''(\tau_2)\right)G(Q_1,\tau_1,Q_2,\tau_2)=0
\label{edp}
\end{equation}
with $G(Q_1,\tau,Q_2,\tau)=\delta(Q_1-Q_2)$ and $
G(0,\tau_1,Q_2,\tau_2)=G(Q_1,\tau_1,0,\tau_2)=0$.
In our problem $f''(\tau)=2$.

The equation (\ref{edp}) can be solved 
(see \cite{fmp} for details) and one finds 
\begin{equation}
G(Q_1,\tau_1,Q_2,\tau_2)=\sum_{k\geq 1} {\rm e}^{-\omega_k(\tau_2-\tau_1)}
{{\rm Ai}(Q_1-\omega_k){\rm Ai}(Q_2-\omega_k)\over({\rm Ai}'(-\omega_k))^2}
\label{sol}
\end{equation}
where ${\rm Ai}$ is the Airy function\cite{as} which has an infinite 
countable numbers of zeroes $-\omega_k$ on the negative real axe
($-\omega_1= -2.33811\ldots, -\omega_2 -4.08795\ldots, \ldots)$.
This function had to be expected in this problem as it 
is known that it arises in the description of 
a Brownian motion with a parabolic 
drift \cite{gr}.

Using Eqs.(\ref{ik},\ref{gg},\ref{sol}), one gets 
\begin{equation}
\bar I_m(M)={m\over \pi}{\rm e}^{-M^3/12}{\cal I}(M)+{\cal O}(m^{2})
\end{equation}
with 
\begin{equation}
{\cal I}(M)=\sum_{k\geq 1} {\rm e}^{-\omega_k M}.
\end{equation}

In the same way, I use Eqs.(\ref{jk},\ref{gg},\ref{sol}) and obtain
\begin{equation}
\bar J_m(Y)=\sqrt{{m\over\pi}}\lim_{{\cal M}\rightarrow \infty}
{\rm e}^{(Y/2-{\cal M})^3/3-(Y/2)^3/3}
\int_0^\infty dx\, {\rm e}^{-x(Y/2-{\cal M})} 
\sum_{k\geq 1}{\rm e}^{-\omega_k{\cal M}}{{\rm Ai}(x-\omega_k)
\over {\rm Ai}'(-\omega_k)}+{\cal O}(m^{3/2}).\label{jj}
\end{equation}
Using the integral representation of the sum in (\ref{jj})\cite{gr}, 
\begin{equation}
\sum_{k\geq 1}{\rm e}^{-\omega_k{\cal M}}{{\rm Ai}(x-\omega_k)
\over {\rm Ai}'(-\omega_k)}=\int_{c-i\infty}^{c+i\infty}dz\,
{\rm e}^{z{\cal M}}{{\rm Ai}(z+x)
\over {\rm Ai}(z)}
\end{equation}
with $c>-\omega_1$,
one can exchange the integration and find after a 
tedious analytical calculation\cite{fmp}
\begin{equation}
\bar J_m(Y)=\sqrt{{m\over\pi}}{\rm e}^{-(Y/2)^3/3}{\cal J}(Y)+{\cal O}(m^{3/2})
\end{equation}
with
\begin{equation}
{\cal J}(Y)={1\over 2\pi i}\int_{c-i\infty}^
{c+i\infty}dz {{\rm e}^{zY/2}\over {\rm Ai}(z)}
\end{equation}
where $c>-\omega_1$.

Now, inserting the expression for $I$ and $J$ in Eq.(\ref{rhoij}), one has
in the original variables
\begin{equation}
\rho(M,P;t)={m^2\over t^{5/3} \pi^2}{\cal I}\left({M\over t^{2/3}}\right)
{\cal J}\left(-{M\over t^{2/3}}+{Pt^{1/3}\over M}\right)
{\cal J}\left(-{M\over t^{2/3}}-{Pt^{1/3}\over M}\right)
+{\cal o}\left({m^2\over t^{5/3}}\right)
\end{equation}

From this equation one has that the concentration $c(t)$ of aggregates, 
the aggregates average mass and momentum and the mean energy per unit 
of length $E(t)$ behave, for time $t\gg 1$, as
\begin{equation}
c(t)\sim t^{-2/3},\quad \langle M\rangle_t\sim t^{2/3},\quad 
\langle P\rangle_t=0,\quad \sqrt{\langle P^2\rangle_t}\sim t^{1/3},\quad
E(t)\sim t^{-2/3}.
\end{equation}

A careful integration over $P$ \cite{fmp} leads to the mass distribution,
which is the main result of this letter,
\begin{equation}
\rho(M,t)={1\over t^{4/3}}F\left({M\over t^{2/3}}\right)
+{\cal o}\left({m^2\over t^{4/3}}\right)
\label{rho}
\end{equation}
where one sees that it obeys the expected scaling form with a scaling function
\begin{equation}
F(M')=2{m^2\over \pi^2}M'{\cal I}(M'){\cal H}(M')
\end{equation}
where
\begin{equation}
{\cal I}(M')=\sum_{k\geq 1} {\mathrm e}^{-\omega_k M'},\quad
{\cal H}(M')={1\over 2\pi i}\int_{c-i \infty}^{c+i\infty}dz\,
{{\rm e}^{-M'z}\over {\rm Ai}^2(z)}
\end{equation}
with $c>-\omega_1$.

\begin{figure}
\narrowtext
\epsfxsize=\hsize
\epsfbox{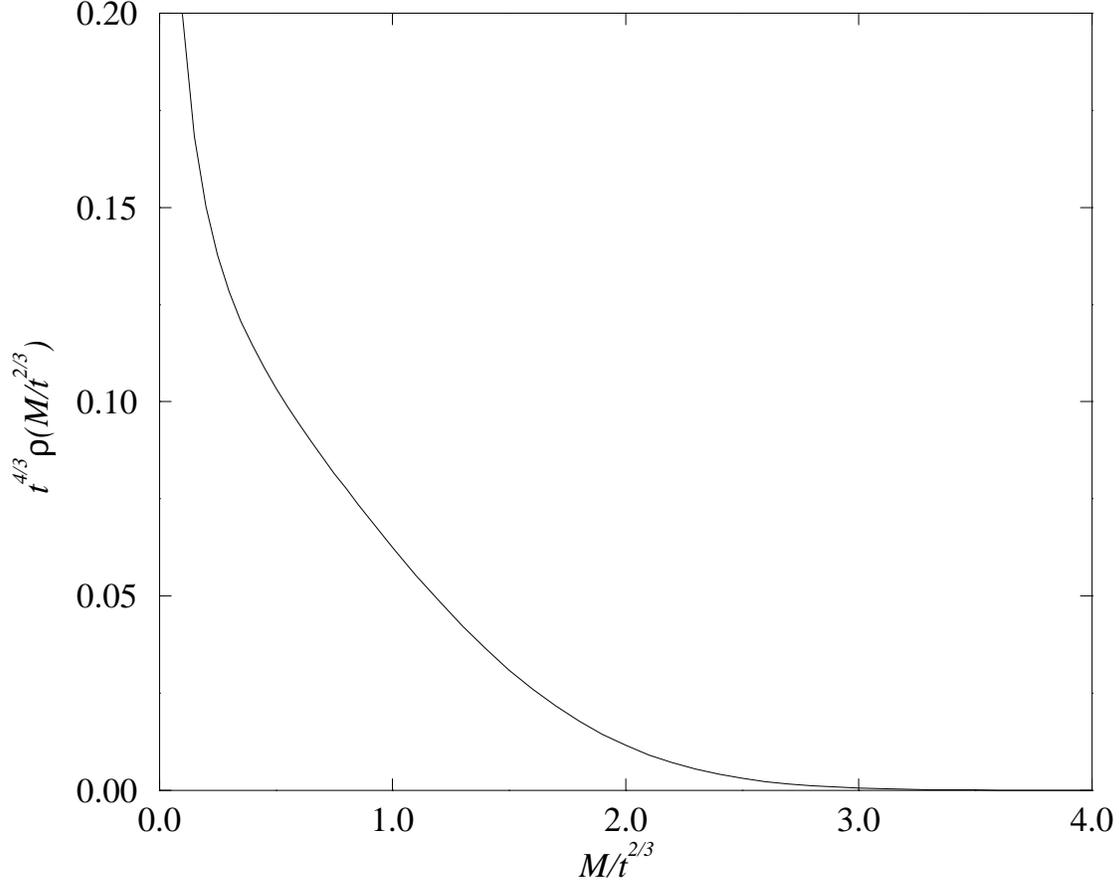}
\caption{
The rescaled mass distribution $t^{4/3}\rho(M,t)$ 
as a function of $M/t^{2/3}$.
\label{fig2}}
\end{figure}

The scaling function $F(M')$ is plotted on Fig.\ref{fig2}.

One can compare the obtained scaling function with the conjectured 
one ($F_{\mathrm conj.}(M')=\exp(-M')$)\cite{cpy}. 
In particular small and large arguments present 
strong differences.
Indeed, for $M'\ll 1$, 
one get ${\cal H}(M')=1+{\cal O}(M')$ while one can estimate 
${\cal I}(M')$ using the asymptotic properties of the zeroes of the 
Airy function $\omega_k=[(3\pi k)/2]^{2/3}+{\cal O}(k^{-1/3})$ and find
${\cal I}(M')\sim (2\sqrt{\pi}M'^{3/2})^{-1}$. One thus find
\begin{equation}
F(M')={m^2\over \pi^{5/2}}{1\over \sqrt{M'}}+{\cal O}(\sqrt{M'}),
\quad (M'\ll 1).
\end{equation}
One can conclude, for example, 
that the number $N(M_0,t)=\int_0^{M_0}\rho(M,t)
\,dM$ of aggregates of small masses $M<M_0\ll t^{2/3}$ 
at time $t$ is well 
underestimated by the conjectured form which leads to $N(M_0,t)\sim M_0/t$, 
while the exact solution gives $N(M_0,t)\sim \sqrt{M_0}/t$. 

For $M'\gg 1$, one can estimate the function ${\cal H}(M')$ by the
steepest descent method and find ${\cal H}(M')\sim
\sqrt{\pi}M'^{3/2}\exp(-M'^3/12)$.
On the other end, one has ${\cal I}(M')\sim\exp(-\omega_1 z)$ and finally
\begin{equation}
F(M')={2\over\pi^{3/2}}M'^{5/2}{\mathrm e}^{-M'\omega_1-M'^3/12},
\quad (M'\gg 1).
\end{equation}
This is again different from the conjectured function as large masses 
$M\gg t^{2/3}$ have a much smaller chance to be present in the system.

Notice that the asymptotic behaviors of the scaling function 
are compatible with the exact
bounds found for the burgers problem\cite{avellaneda} 
with white noise initial condition. 
On the other end, Eq.(\ref{rho}) solves
the shock strength distribution questioned in \cite{burgers} and studied
numerically in \cite{kida}.

It is instructive to compute, along the same line, the collision frequency
$\nu_2(M_1,M_2,t)$ 
where $\nu_2\,dM_1\,dM_2\,dt$ 
is the number of collision per unit of volume between
masses in $(M_1,M_1+dM_1)$ 
and $(M_2,M_2+dM_2)$ in a time in $(t,t+dt)$.
I find 
\begin{eqnarray}
&&\nu_2(M_1,M_2,t)=\int_{-\infty}^{\infty}dP_1\int_{-\infty}^{\infty}dP_2\,
\left|{P_1\over M_1}-{P_2\over M_2}\right|
\delta\left(t\left({P_1\over M_1}-{P_2\over M_2}\right)-M_1-M_2\right)
\nonumber\cr\\
&&\quad\times t^{-2/3}J_{m'}\left(-{P'_1\over M'_1}-M'_1\right)
I_{m'}(M'_1,P'_1)I_{m'}(M'_2,P'_2)
J_{m'}\left({P'_2\over M'_2}-M'_2\right),
\end{eqnarray}
with $M_i'=M_i/t^{2/3}$, $P_i'=P_i/t^{1/3}$ and $m'=m/t^{2/3}$,
leading to
\begin{equation}
\nu_2(M_1,M_2,t)\sim\left({m\over t\pi}\right)^3 
{\cal F}\left({M_1\over t^{2/3}},{M_2\over t^{2/3}}\right)
\end{equation}
with
\begin{equation}
{\cal F}(M_1',M_2')=(M_1'+M_2')
M'_1M_2'{\cal I}(M_1')
{\cal I}(M_2'){\cal H}(M_1'+M_2')
\end{equation}
and ${\cal I}$ and ${\cal H}$ as above.
This collision frequency clearly does not factorize in a product of functions
of $M_1$ and $M_2$, respectively. 
This fact invalidates the assumption on which
the mass distribution was computed in \cite{p}.

One can inquire about the universality of these results with respect to other 
initial conditions. Let us first consider a Poissonian distribution of 
the particles initial positions with an average interparticle distance $a$.
The discrete points over which the Brownian motion
should pass in the construction of our solution are still distributed 
on the same parabola but with irregular spacing. 
In the long time limit and after rescaling, 
the spacing between points of average $a'=a/t^{2/3}$ become smaller and smaller
up to be, in first order in $m'$, a continuum. 
The difference between irregular and regular spacing is thus asymptotically 
erased and the result Eq.(\ref{rho}) should be recovered in this case.

A bimodal momentum distribution $\phi(p)=(\delta(p-p_0)+\delta(p+p_0))/2$ 
is used in the initial state in \cite{cpy}. I believe that this 
should not affect the form of the mass distribution (\ref{rho})
as the random walk initiated by this distribution is 
well approximated, in the long time limit, by the considered Brownian motion.

One can define a distribution where momentum are initially correlated. 
In this case, one expects the scaling function to be different, 
at least for small $M'$\cite{sinai}.

In summary, I found an exact asymptotic solution for the mass distribution
of the ballistic aggregation in one dimension. 
Such an exact solution is not frequent in a nonequilibrium system and has 
permitted to verify scaling hypothesis for this system. 
While the average mass per aggregate
was proved to behave with time as $\langle M\rangle_t\sim t^{2/3}$ 
for $t\gg 1$, as expected from previous studies, the scaling function 
is shown here to be different from the conjectured one.
This distribution also solve the shock 
strength distribution of the one dimensional Burgers equation in the 
inviscid limit with a white noise initial condition. 

I gratefully acknowledge numerous useful discussions with P. Martin and 
J. Piasecki and financial support from the Swiss National Foundation.

\end{document}